\documentclass[12pt]{iopart}
\usepackage{iopams}

\usepackage{graphicx}

\newcommand\be{\begin{equation}}
\newcommand\bea{\begin{eqnarray}}
\newcommand\eea{\end{eqnarray}}
\newcommand\ee{\end{equation}}

\newcommand{\D}{\mathrm{d}}

\newcommand{\vecr}{(\mathbf{r})}
\newcommand{\hn}{\hat{n}}

\newcommand{\kbt}{k_{\mathrm{B}}T}

\newcommand{\lb}{l_\mathrm{B}}
\newcommand{\ld}{\lambda_\mathrm{D}}
\newcommand{\lgc}{\lambda_\mathrm{GC}}

\newcommand{\phia}{\phi_{\mathrm{A}}}
\newcommand{\phib}{\phi_{\mathrm{B}}}
\newcommand{\epsa}{\varepsilon_{\mathrm{A}}}
\newcommand{\epsb}{\varepsilon_{\mathrm{B}}}
\newcommand{\epsz}{\varepsilon_{0}}
\newcommand{\epsr}{\varepsilon_{\mathrm{r}}}

\begin{document}

\title[Ion-specific interactions in aqueous solutions]
{Beyond standard Poisson-Boltzmann theory: ion-specific interactions in aqueous solutions}

\author{Dan Ben-Yaakov$^1$, David Andelman$^1$, Daniel Harries$^2$ and Rudi Podgornik$^{3}$}
\address{$^1$ Raymond and Beverly Sackler School of Physics and Astronomy, Tel Aviv
University, Ramat Aviv, Tel Aviv 69978, Israel}
\address{$^2$ Institute of Chemistry and The Fritz Haber Research Center, The Hebrew
University, Jerusalem 91904, Israel}
\address{$^3$ Department of Theoretical Physics, J. Stefan
Institute, Department of Physics, Faculty of Mathematics and Physics and
Institute of Biophysics, Medical Faculty, University of Ljubljana, 1000 Ljubljana, Slovenia}
\eads{\mailto{benyaa@post.tau.ac.il}, \mailto{andelman@post.tau.ac.il}, \mailto{daniel@fh.huji.ac.il},
\mailto{rudolf.podgornik@fmf.uni-lj.si}}

\begin{abstract}
The Poisson-Boltzmann mean-field description of ionic solutions has been successfully
used in predicting charge distributions and interactions between charged macromolecules.
While the electrostatic model of  charged fluids, on which the Poisson-Boltzmann
description rests, and its statistical mechanical consequences have been scrutinized
in great detail, much less is understood about its probable shortcomings when dealing
with various aspects of real physical, chemical and biological systems. These
shortcomings are not only a consequence of  the limitations of the mean-field
approximation {\em per se}, but perhaps are primarily due to the fact that the
purely Coulombic model Hamiltonian does not take into account various additional
interactions that are not electrostatic in their origin. We explore several
possible non-electrostatic contributions to the free energy of ions in confined
aqueous solutions and investigate their ramifications and consequences on ionic
profiles and interactions between charged surfaces and macromolecules.
\end{abstract}

\pacs{61.20.Qg,87.15.N-, 82.60.Lf}
\submitto{\JPCM}
\maketitle

\section{Introduction}

The traditional approach to ions in solution has been the mean-field Poisson-Boltzmann
(PB) formalism. This approach adequately captures the main features of electrostatic
interactions at weak surface charges, low ion valency, and high temperature
\cite{Andelman,Oosawa}. It stems from a {\em Coulombic} model Hamiltonian that includes
only purely electrostatic interactions between different charged species. The
limitations of the PB approach, which rests on a collective and continuous
description of statistical charge distributions become particularly important
in highly-charged systems, where counterion-mediated interactions between
charged bodies cannot be described by the mean-field approach that completely
neglects ion correlations and charge fluctuations \cite{Naji}. These mean-field
limitations have been successfully bypassed and have led to  more refined
descriptions that capture some of the important non-mean-field aspects of Coulomb fluids \cite{hoda}.

Beside  electrostatic interactions that are universal, omnipresent
non-electrostatic interactions are
specific and dependent on the nature of the ionic species,
solvent and  confining interfaces.
Because it is difficult, perhaps impossible, to devise a universal
theory accounting for all non-electrostatic effects,  such additional
interactions need to be treated separately, describing specific electrolyte
features that go beyond regular PB theory. These effects depend on ionic chemical nature, size,
charge, polarizability and  solvation (preferential ion-solvent interaction \cite{marcus}).

In this review we describe how adding non-electrostatic terms in the system free energy
yields modified PB equations and ionic density profiles and
differentiate between the way different ionic species interact with
macromolecules. While it has long been recognized that ions
can have such different effects on macromolecular interactions \cite{cviklik},
only recently have such ionic features been accounted for
within electrostatic mean-field theory.
Examples include ion effects on surface tension \cite{onuki}
and precipitation of proteins from solutions \cite{currentninham}. The latter led
to the so-called {\em Hofmeister ranking} of different ions according to
their surface activity. Adsorption of ions
and/or surfactants to charged interfaces also contains substantial
contributions from non-electrostatic degrees of freedom,  as
was shown within the Ninham-Parsegian theory of charge regulation at surfaces
\cite{chargeregulation}.

The outline of this paper is as follow. After reviewing
the standard PB theory in section 2, we present in section 3 how addition of
steric effects  result in
saturation of ionic profiles close to charged interfaces. In section 4 we show
how non-electrostatic  interactions between ions and charged membranes can cause
a phase transition between two  lamellar systems of different periodicity.
Solvation effects are the topic discussed in section 5, where local variation
of the dielectric function in solvent mixtures and ion-solvent interactions lead
to changes in ionic and solvent profiles close to charged interfaces.
Finally, in section 6 possible polarization effects of ions in solution is added
to their ionic character, again resulting in different behavior close to charge interfaces.

\section{The Poisson-Boltzmann model: summary and main results}

\begin{figure}\centering
\includegraphics[width=80mm]{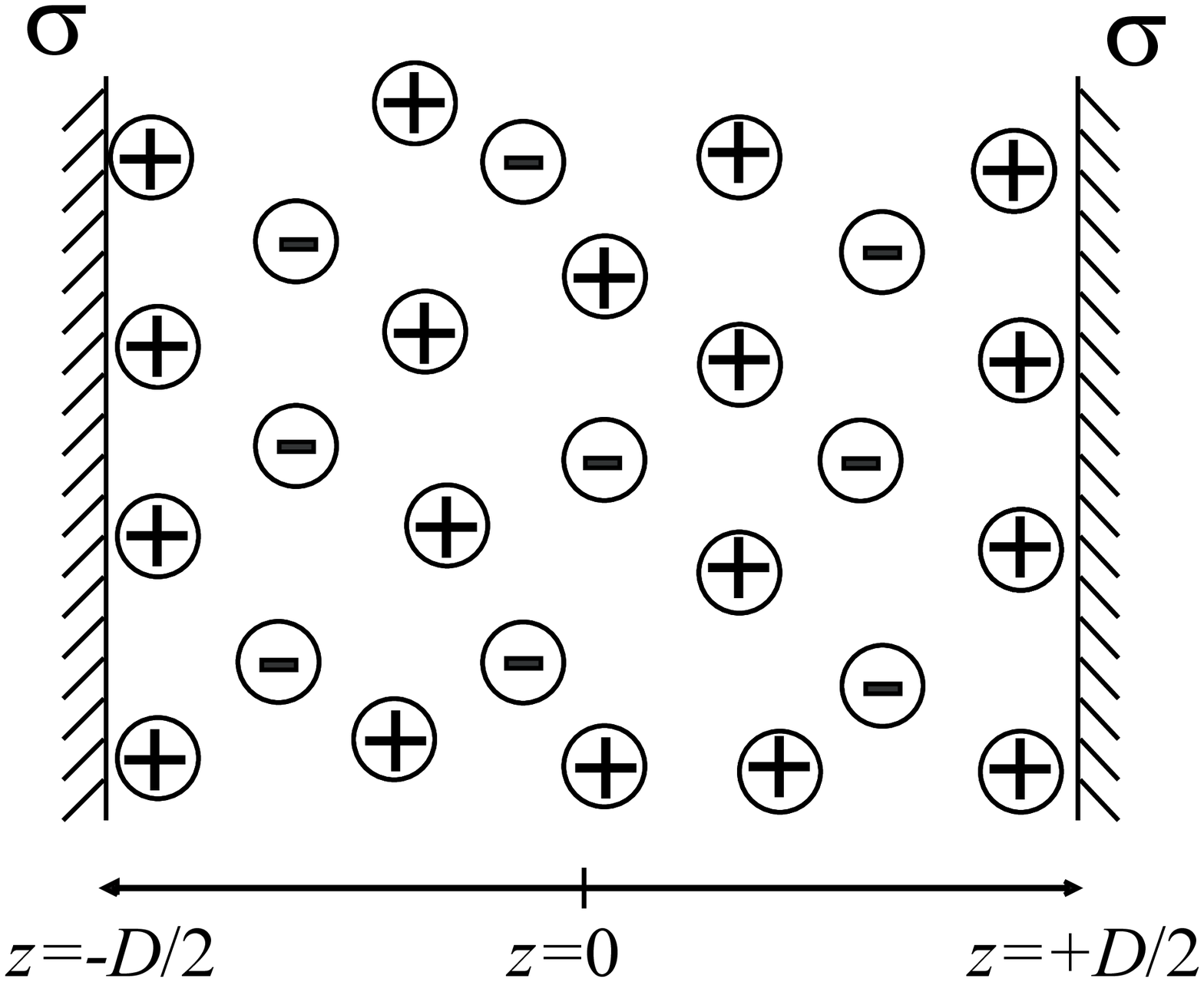}
\caption{\footnotesize\textsf{Schematic illustration of the model system. The two
plates residing at $z=\pm D/2$ are charged with surface charge
density $\sigma$. The electrolyte ions are denoted by $\oplus$ and $\ominus$.
Their densities are $n_{+}(z)$ and $n_{-}(z)$. }} \label{fig1}
\end{figure}

The PB theory  is a  useful starting point for many theoretical ramifications because
it relies on a simple and  analytically tractable model that can be easily extended
and amended. Although the theory has its own well-understood limitations,
it yields meaningful results in good
agreement with many experiments \cite{Andelman}.

The model can be cast in many geometries but we shall focus on the simplest
planar system, as depicted in figure~\ref{fig1}. Two charged planar
plates (of infinite extent), located at  $z=\pm D/2$, are immersed in an
electrolyte bath. Each plate carries a charge density of $\sigma$ (per unit
area) and their inter-surface separation $D$ is a tunable system parameter. The ionic
solution is denoted as $z_{+}{:}z_{-}$ and is composed of a solvent of
dielectric constant $\varepsilon$ and two types of ions:  cations of valency
$z_{+}$ and local density (number per unit volume) $n_{+}(z)$, and anions of
valency $z_{-}$ with local density $n_{-}(z)$. The finite system of thickness
$D$ is in contact with an electrolyte bath of bulk densities $n_{\pm}^b$ obeying
the electroneutrality condition: $z_{+}n^b_{+}=z_{-}n^b_{-}$.
Throughout the paper, we will use the
convention that $z_{-}$ (denotes the valency of the anions) is
taken as a positive number, hence, their respective charge  is written as ${-}ez_{-}$.

The free energy within the PB model $F_{\rm PB}$ can be derived
in numerous ways as can be found in the literature
\cite{Andelman}. With our notation $F_{\rm PB}$ is a function of the densities $n_{\pm}$
and the local electrostatic potential $\psi$:

\bea
\label{F_PB}
 F_{\rm PB}&=&\int \D^3r \left[-\frac{\varepsilon}{8\pi}(\nabla\psi)^2 +
 (z_{+}n_{+}-z_{-}n_{-})e\psi\right] \nonumber\\
&+& \kbt \int \D^3r \left[n_{+}\ln(a^3n_{+})+ n_{-}\ln(a^3n_{-}) -n_{+}-n_{-}\right]\nonumber\\
&-& \int \D^3r \left[\mu_{+}n_{+} +\mu_{-}n_{-}\right]
\eea
The above free energy is a functional of three independent
fields: $n_{\pm}$ and $\psi$. The first integral in (\ref{F_PB}) is the
electrostatic energy, while the second represents the ideal mixing
entropy of a dilute solution of the $\pm$ ions. Note that a
microscopic length scale $a$ was introduced in the entropy terms above.
Only one such microscopic length scale will be used throughout this
paper, defining a reference density associated with close-packing
${n_0}=1/a^3$.

The last integral in (\ref{F_PB}) is written in
terms of the chemical potentials $\mu_{\pm}$ of the $\pm$ ions. Alternatively, the chemical potential
can be regarded as a Lagrange multiplier, setting the bulk densities
to be $n_{\pm}^b=\exp(\mu_{\pm}/\kbt)/a^3$.

The thermodynamic equilibrium state is given by minimizing the above free energy
functional. Taking the variation of the free energy (\ref{F_PB}) with respect
to $n_\pm$ yields the Boltzmann distribution of the ions in the presence of the local potential $\psi$:
\bea
\frac{\delta F_{\rm PB}}{\delta n_{\pm}}&=& \pm e z_{\pm}\psi+\kbt\ln(n_{\pm}a^3)-\mu_{\pm}=0 \label{bol_dist}
\eea
wherefrom
\be
n_{\pm}=n_{\pm}^b\exp\left(\mp ez_{\pm}\psi/\kbt\right)
\ee
Similarly, taking the variation with respect to the potential $\psi$ yields
the Poisson equation connecting $\psi$ with the $\pm$ densities:
\bea
\frac{\delta F_{\rm PB}}{\delta \psi}&=&\frac{\varepsilon}{4\pi}
\nabla^2\psi + ez_{+}n_{+}-ez_{-}n_{-}=0
\label{pos_eq}
\eea
Combining the Boltzmann distribution with the Poisson equation yields
the familiar Poisson-Boltzmann equation:
\be
\nabla^2\psi=-\frac{4\pi e}{\varepsilon}
\left[ z_{+}n_{+}^b\e^{-ez_{+}\psi/\kbt}\,-\, z_{-}n_{-}^b \e^{ez_{-}\psi/\kbt}\,\right]
\ee
that serves as a starting point for various extensions presented in the sections to follow.

In addition to the volume contribution (\ref{F_PB}) of the free energy, we need to include a
surface electrostatic energy term $F_s$, which couples the surface charge density $\sigma$ with
the surface value of the potential $\psi(z{=}\pm D/2)=\psi_s$. This surface term has the form
\be
\label{Fs}
F_s=\int_{\rm A} \D^2r \, \sigma\psi_s
\ee
and takes into account the fact that we work in an ensemble
where the surface charge density is fixed. Variation of $F_{\rm PB}
+F_s$ with respect to $\psi_s$ yields the well-known electrostatic boundary condition
\bea
\frac{\delta }{\delta \psi_s}(F_{\rm PB}+F_s) = \frac{\varepsilon}{4\pi}
\mathbf{\hat{n}}\cdot{\nabla}\psi\, +\, \sigma=0
\eea
wherefrom
\bea
\mathbf{\hat{n}}\cdot{\nabla}\psi=-\frac{4\pi}{\varepsilon}\sigma
\eea
where $\mathbf{\hat{n}}$ is a unit vector normal to the surface.
This boundary condition can also be interpreted as the electroneutrality
condition since the amount of mobile charge should exactly compensate the surface charge.

The above PB equation is valid in any geometry. But if we go back to the
planar case with two parallel plates, it is straightforward to calculate the local pressure $P_{\rm PB}$ at
any point $z$ between the plates. In chemical equilibrium  $P_{\rm PB}$
should be a constant throughout the system. Namely, $P_{\rm PB}$ is independent of the position $z$
and can be calculated by
taking the proper variation of the free energy with respect to the
inter-plate spacing $D$:
\be
P_{\rm PB}=-\frac{\varepsilon}{8\pi}(\psi^\prime)^2 +\kbt\left(n_{+} + n_{-}\right)
\label{P_PB}
\ee
The  pressure is composed of two terms. The first is the electrostatic pressure
stemming from the Maxwell stress tensor. This term is negative, meaning an
attractive force contribution acting between the plates. The second term
originates from the ideal entropy of mixing of the ions and is positive.
This term is similar to an ``ideal gas'' van 't Hoff osmotic pressure of the $\pm$ species.

A related relation can be obtained from (\ref{P_PB}) for one charged surface.
Comparing the pressure $P_{\rm PB}$ calculated at contact with the charge surface,
with the distal pressure ($z\to \infty$) results in the
so-called Graham equation used in colloid and interfacial science \cite{evans}
\be
\label{graham_PB}
\sigma^2= \frac{\varepsilon\kbt}{2\pi}(n_{+}^s+n_{-}^s-n_{+}^b-n_{-}^b)
\ee
where $n_{\pm}^s$ are the values of the counterion and co-ion densities calculated at the surface.

Note that the osmotic pressure as measured in experiments can be written as the
difference between the local pressure and the electrolyte bath pressure:
\be
\Pi=P_{\rm PB}-\kbt\left(n_{+}^b+n_{-}^b\right)
\ee

We now mention two special cases separately: the {\em counterion only} case and the
linearized Debye-H\"uckel theory.

\subsection{The counterion only case}

In the {\em counterion only} case
no salt is added to the solution and there are just enough counterions to balance the surface charge.
For this case we  have chosen arbitrarily the sign of the surface charge to be negative, $\sigma<0$. By
setting  $n_{+}\equiv 0$, the free energy is written only in terms
of the anion density in solution, $n\equiv n_{-}$ and $z_{-}=z$
\be\label{F_PB_counter}
F_{\rm PB}=\int \D^3r \left[-\frac{\varepsilon}{8\pi}(\nabla\psi)^2 - e z n\psi
+ \kbt (n\ln\frac{n}{n_0} -n)\right]
\ee
The local pressure $P_{\rm PB}$ also contains only osmotic pressure of one type
of ions ($n_{-}$), apart from the Maxwell stress term.

\subsection{The linearized Debye-H\"uckel theory}

When the surface charges and potentials are small, $e\psi_s\ll
\kbt$ ($\psi_s\ll 25$\,mV  at room temperature), the PB equation can be
linearized and matches the Debye-H\"uckel theory. For the simple and symmetric
$1{:}1$ monovalent electrolytes the PB equation reduces to:
\be
\nabla^2 \Psi=\ld^{-2}\Psi
\ee
where the dimensionless potential is defined as $\Psi=e\psi/\kbt$ and the Debye screening length is
$\ld=\sqrt{\varepsilon\kbt/8\pi e^2 n_b}$. The main importance of the Debye
length $\ld$ is to indicate a typical length for the exponential decay
of the potential around charged objects and boundaries:
$\Psi(z)\sim \exp(-z/\ld)$. The $\ld$ length is about 3\AA\ for $n_b=1$\,M of NaCl and about
1$\mu$m for pure water. It is convenient to express the Debye length in
term of the Bjerrum length $\lb=e^2/\varepsilon\kbt$ as $\ld=\sqrt{1/
8\pi \lb n_b}$. For water at room temperature $\lb\simeq 7$\AA.

\bigskip

Returning now to the general non-linear PB model,
its free energy (\ref{F_PB})
can also be expressed
in terms of the dimensionless potential $\Psi$.  For a 1:1 symmetric
electrolyte we write it as:
\bea
\label{F_PB_dimensionless}
F_{\rm PB}/\kbt&=&\int \D^3r \left[-\frac{1}{8\pi\lb}(\nabla\Psi)^2 +(n_{+}-n_{-})\Psi\right] \nonumber\\
&+&  \int \D^3r \left[n_{+}\ln(n_{+}/n_b)+ n_{-}\ln(n_{-}/n_b)  \right.  \nonumber\\
&- & \left.  n_{+}-n_{-}+2n_b\right]
\eea
where the bulk contribution to the free energy is subtracted in the above expression.

In the next sections we will elaborate on several extensions
and modifications of the PB treatment. Results in terms of ion
profiles and inter-surface pressure will be presented and compared to the bare PB results.

\section{Steric effects: finite ion size}

At sufficiently high ionic densities steric effects prevent
ions from accumulating at charged interfaces to the extent predicted by the standard
PB theory. This effect has been noted already in the work of
Eigen \cite{eigen}, elaborated later in \cite{iglic} and
developed into a final form by Borukhov et al. \cite{Borukhov} and more recently in \cite{tresset}.
Steric constraints lead to saturation of ion density
near the interface and, thus, increase their concentration
in the rest of the interfacial region. This follows quite generally
from the energy-entropy competition: the gain from the
electrostatic energy is counteracted by the entropic penalty
associated with ion packing.
%
Beyond the mean-field  ({\em e.g.}, integral equation closure
approximations), the correlations in local molecular packing clearly lead to
ion layering and  non-monotonic interactions between interfaces \cite{israelachvili,messina}.

Steric effects lead to a modified ionic entropy that in turn
gives rise to a modified Poisson--Boltzmann (MPB) equation,
governing the distribution of ions in the vicinity of charged
interfaces. The main features of the steric effect can be derived from a lattice-gas model
introduced next \cite{hill}. Other possible approaches that were
considered include the Stern layer
modification of the PB approach, extensive MC simulations
or numeric solutions of the integral closures relations \cite{pbapproach}.

We start with a free energy where the entropy of mixing is taken in
its exact lattice-gas form, without taking the dilute solution limit.
Instead of point-like particles, the co-ions and counter-ions
are now modeled as finite-size
particles having the same radius $a$.
The free energy for a $z_{+}{:}z_{-}$ electrolyte is now a modification of $F_{\rm PB}$ (\ref{F_PB}):
\bea
\label{MPB_free_energy}
 F_{\rm MPB} &= &\int \D^3r \left[-\frac{\varepsilon}
{8\pi}(\nabla\psi)^2 +(z_{+}n_{+}-z_{-}n_{-})e\psi\right] \nonumber\\
& & + \kbt \int \D^3r \left[n_{+}\ln(a^3n_{+})+ n_{-}\ln(a^3n_{-})\right]\nonumber\\
& & + \frac{\kbt}{a^3} \int \D^3r\, (1-a^3n_{+}-a^3n_{-})\ln(1-a^3n_{+}-a^3n_{-}) \nonumber\\
&&  - \int \D^3r \left[\mu_{+}n_{+} +\mu_{-}n_{-}\right]
\eea

Taking the variation with respect to the three fields:
$n_{\pm}$ and $\psi$, yields the MPB equilibrium equations.
We give them below for two cases: (i)  symmetric
electrolytes $z=z_{+}=z_{-}$; and, (ii)  $1{:}z$ asymmetric ones.

In the former case the MPB equation is written as:
\bea
\nabla^2\psi&=&-\frac{8\pi ez n_b}{\varepsilon}\frac{\sinh(ez\psi/\kbt)}
{1-\hn + \hn\cosh(ez\psi/\kbt)}
\eea
while the local ion densities are
\bea
n_\pm&=&n_b\frac{\exp(\mp e z \psi/\kbt)}{1-\hn+\hn\cosh(ez\psi/\kbt)}
\label{anotherone-1}
\eea
where $\hn=a^3(n_+^{\rm b}+n_-^{\rm b})$ is the  bulk volume fraction of the ions.
Clearly, the ion densities saturate for large values of the electrostatic
potential, preventing them from reaching unphysical values that can be
obtained in the standard PB theory.

In the latter case of 1:$z$ electrolytes, the MPB is obtained in the form

\bea
\nabla^2\psi &=& -\frac{4\pi}{\varepsilon}(en_{+}-ezn_{-})
\eea
with the corresponding local densities
\numparts
\bea
n_{+}&=&\frac{zn_b\e^{-e\psi/\kbt}}
{1-\hn+\hn\left(\e^{ez\psi/\kbt}+z\e^{-e\psi/\kbt}\right)/(1+z)} \\
n_{-}&=& \frac{n_b\e^{ze\psi/\kbt}}
{1-\hn+\hn\left(\e^{ez\psi/\kbt}+z\e^{-e\psi/\kbt}\right)/(1+z)}
\eea
\endnumparts
As discussed above, for large electrostatic potentials the densities saturate at
finite values dependent on $\hn$, $z$ and $n_b$.

\begin{figure}\centering
\includegraphics[width=70mm]{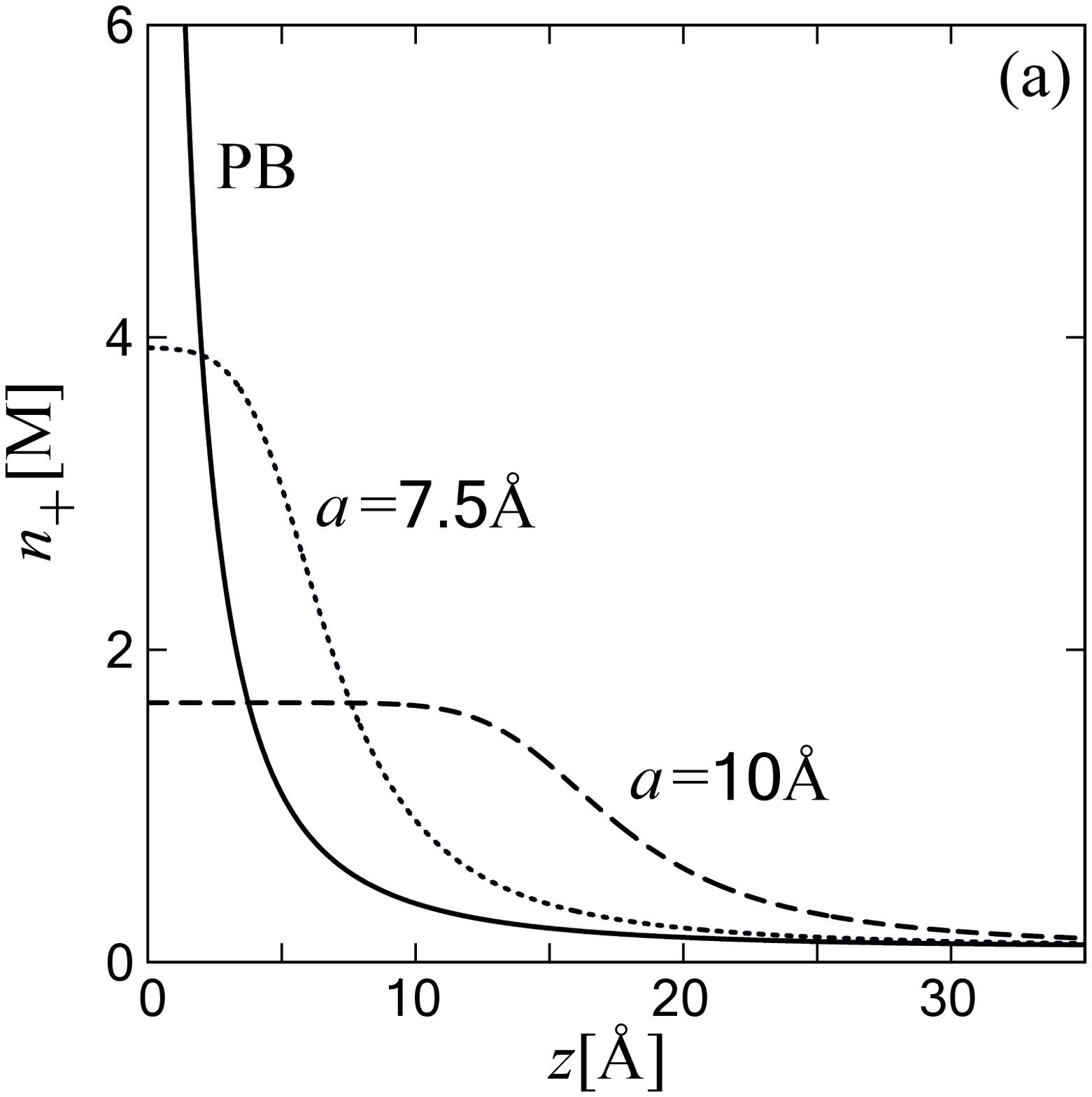}
\includegraphics[width=70mm]{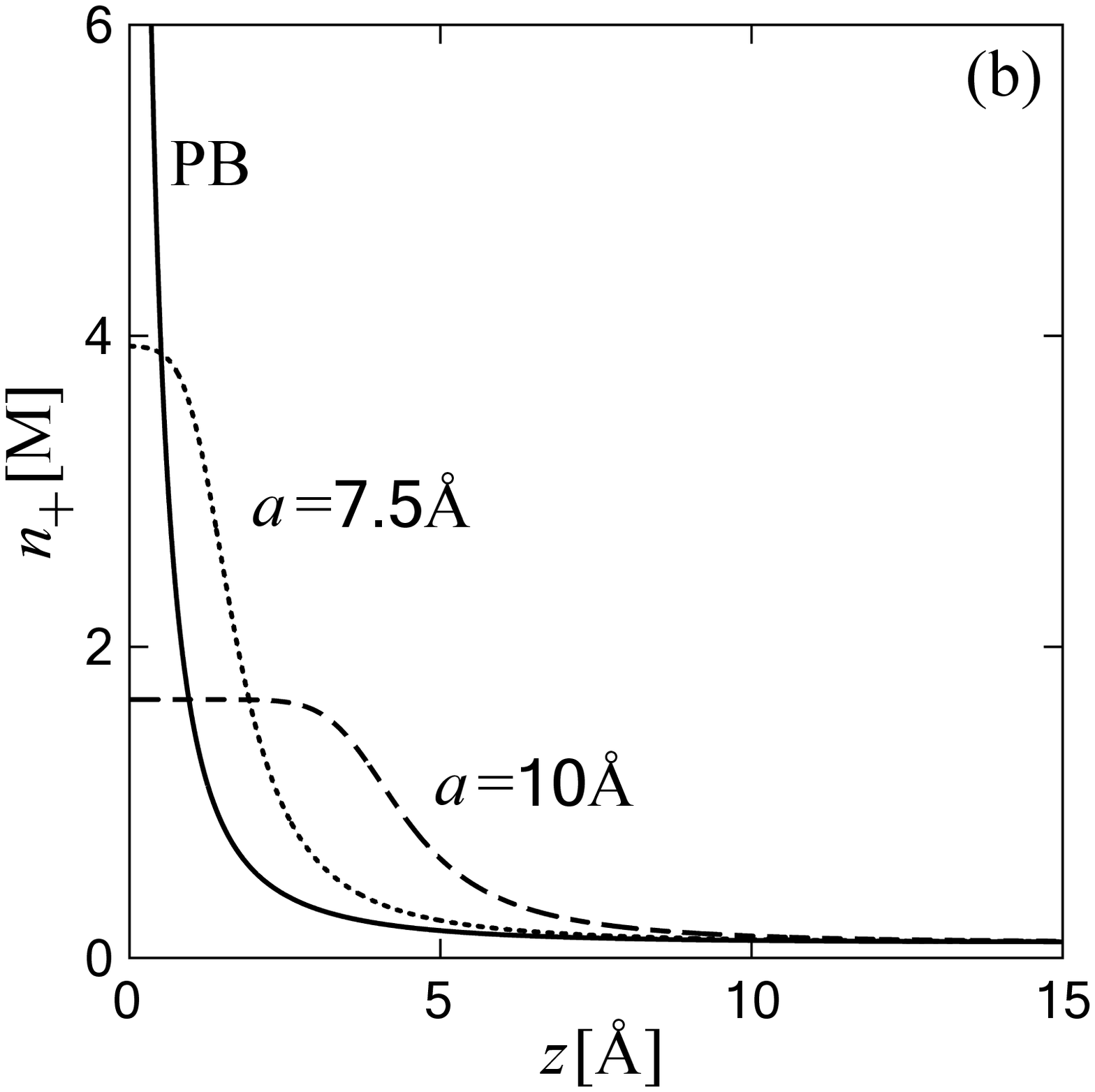}
\caption{\footnotesize\textsf{Counterions density profile
calculated for the MPB model. The solid line represents the
standard PB model, while the dotted and dashed lines
represent the MPB results with molecular sizes of
$7.5$\AA\ and $10$\AA, respectively. In (a) the salt
is symmetric $1{:}1$, while in (b) it is asymmetric $1{:}4$.
Other parameters are: $\sigma=-1/50\,$\AA$^{-2}$, $n_b=0.1$M ,
and $\varepsilon=80$.}}\label{fig2}
\end{figure}

Figure \ref{fig2} shows  the ion density profile close to a single
charged interface with fixed surface charge density $\sigma$, in contact with an electrolyte bath
of ionic density $n_{b}$ and for several ionic sizes $a$.
As is clear from figure \ref{fig2} the steric constraints limit
the highest possible density in the vicinity of the charged surface
and, thus, extend the electrostatic double layer further into the
bulk, if compared to the standard PB theory. The extent of the double layer depends
crucially on the hardcore radius of the ions, $a$. Furthermore, the valency of the
counterions also affects the width of the saturated layer, as clearly demonstrated
by the comparison between figure \ref{fig2}a and figure \ref{fig2}b.

The local pressure $P_{\rm MPB}$ can be calculated with analogy to $P_{\rm PB}$ of (\ref{P_PB})
\bea
\label{MPB_pressure}
P_{\rm MPB}&=&-\frac{\varepsilon}{8 \pi}\psi'^2 + \nonumber\\
& & \frac{\kbt}{a^3}\ln\left[1+\frac{1}{z_{+} +
z_{-}}\frac{\hn}{1-\hn}\left(z_{-}
\e^{-e z_{+}\psi/\kbt} + z_{+} \e^{e z_{-} \psi/\kbt}\right)\right]
\eea
and can be cast into the form of the {\em contact theorem}
that relates the value of the equilibrium osmotic pressure to the values of the
surface potential and its first derivative, yielding a generalization of
the standard Graham equation used in the colloid science \cite{evans}
\be
\label{grahame_MPB}
\sigma^2\simeq \frac{\varepsilon\kbt}{2\pi}\frac{1}{a^3}\ln{\frac{1-2a^3n_b}{1-a^3 n_{+}^s}}
\ee
This analytical expression is valid in the limit that the co-ions have a
negligible concentration at
the surface, $n_{-}^s\to 0$. It is instructive to find that for $a\to 0$,
expanding the logarithm in (\ref{grahame_MPB}), we recover the Graham
equation for the standard PB model (\ref{graham_PB}).

\section{PB and charge regulation in lamellar systems}

\subsection{The model free energy and osmotic pressure}

In the previous sections the surface   free energy $F_s$
(\ref{Fs}) was taken in its simplest form assuming a homogeneous
surface charge, {\em i.e.} in the form of a surface electrostatic energy. We consider now a
generalized form of $F_s$ where lateral mixing of charged species is
allowed within the surface and is described on the level of
regular solution theory \cite{hill}.

Experimentally observed lamellar-lamellar phase transitions in charged surfactant
systems \cite{zemb98} provides an example where non-electrostatic,
ionic-specific interactions appear to play a fundamental role. The
non-electrostatic interactions are limited to  charged amphiphilic surfaces
confining the ionic solution. As evidenced in NMR experiments, ions not only
associate differently with the amphiphile-water interface, but their binding
may also restructure the interface they contact \cite{rydall92}. Computer
simulations also indicate that the restructuring of the amphiphilic headgroup
region should be strongly influenced by the size of the counter-ion
\cite{sachs03}. Such conformational changes at the interface are possible
sources of non-ideal amphiphile mixing, because non-electrostatic ion binding
at the interface may effectively create two incompatible types of amphiphiles:
ion-bound and ion-detached.

We proposed a model \cite{etay} based on an extension of the
Poisson--Boltzmann theory to explain the first-order liquid-liquid (${\rm L}_{\alpha}
\rightarrow {\rm L}_{\alpha'}$) phase transition observed in osmotic pressure measurements
of certain charged lamellae-forming amphiphiles \cite{zemb98}.
Our starting point is the same as depicted in figure~\ref{fig1}. The free energy of the confined
ions has several contributions. The volume free energy, $F_{\rm PB}$,  is taken to be the same
as the PB expression for the counterion only case (\ref{F_PB_counter}).
Because all counterions in solution originate from surfactant molecules, their integrated concentration (per unit area)
must be equal in magnitude and opposite in sign to the surface
charge density
\be
2 \sigma=e\int^{D/2}_{-D/2} n(z) \,{d}z\label{e2}
\ee
This is also the electroneutrality  condition and can be translated via
Gauss' law into the electrostatic boundary condition (in Gaussian units):
$\psi'(D/2)=\psi'_s=4\pi\sigma/\varepsilon$, linking the surface
electric field $\psi'_s$ with the surface charge density $\sigma$.

The second part of the total free energy comes from the surface free
energy, $F_s$, of the amphiphiles residing on the planar
bilayers. Here, we deviate from the $F_s$ expression in (\ref{Fs}) because we allow the
surfactants on the interface to partially dissociate their counterion in the spirit of
the Ninham-Parsegian theory of charge regulation \cite{chargeregulation}.
The surface free energy $F_{s}$ has  electrostatic and
non-electrostatic parts as well as a lateral mixing entropy
contribution. Expressed in terms of the dimensionless surface area
fraction $\eta_s=a^2\sigma/e$ of charged surfactants and dimensionless surface potential
$\Psi_s=e\psi_s/\kbt$, the free energy $F_s$ is:
\bea
F_s&=&\frac{\kbt}{a^2}\int_{\rm A}\D^2r\,\left[\eta_s\Psi_s
-{\alpha_s}\eta_s-\frac{1}{2}{\chi_s}\eta_s^2\right.\nonumber\\
~~~~&+& \eta_s\ln\eta_s+(1-\eta_s)\ln(1-\eta_s)\Bigg]
\label{e3}
\eea
The first term couples the surface charge and surface
potential as in (\ref{Fs}), while the additional terms are the enthalpy and entropy of a
two-component  liquid mixture: charged surfactant with area fraction
$\eta_s$ and neutralized, ion-bound surfactants with area fraction
$1-\eta_s$. The dimensionless parameters ${\alpha_s}$ and ${\chi_s}$ are
phenomenological, and denote respectively the
counterion\---surfactant and the surfactant\---surfactant
interactions at the surface. Here, ${\alpha_s}<0$ means that there
is an added non-electrostatic attraction (favorable adsorption free
energy) between counterions and the surface; the more counterions
are associated at the surface, the smaller the amount of remaining
charged surfactant. A positive ${\chi_s}$ parameter represents the
tendency of surfactants on the surface to phase separate into
domains of neutral and charged surfactants.

The total free energy $F_{\rm tot}$ is written
as a functional of the variables $\Psi(z)$, $n(z)$, and a function of $\eta_s$,
and includes the conservation condition, (\ref{e2}), via a
Lagrange multiplier, $\lambda_s$:
\be
 F_{\rm tot}[\Psi,n;\eta_s]=F_{\rm PB}+F_s-
 \lambda_s\left[\eta_s-{a^2}\int_{0}^{D/2}n(z)\,{\rm d}z\right]
\ee
Next, we minimize $F_{\rm tot}$ with respect to the surface variable $\eta_s$, and the two
continuous fields $n(z)$, $\Psi(z)$:\, $\D F_{\rm tot}/\D \eta_s=\delta F_{\rm tot}/\delta n(z)
= \delta F_{\rm tot}/\delta \Psi(z) =0,$ corresponding to three coupled Euler-Lagrange (EL) equations. The first one
connects the surface charge density ${\eta_s}$ with the surface potential $\Psi_s$
\numparts
\bea
\frac{\eta_s}{1-\eta_s}&=&\exp\left(\lambda_s+\alpha_s+\chi_s\eta_s-\Psi_s\right)\label{e5}
\eea
The second one is simply the Boltzmann distribution for the spatially--dependent ion density
\bea
a^3n(z)&=&\exp\left(-\lambda_s+\Psi(z)\right)\label{e6}
\eea
and the last one is the standard Poisson equation
\bea
\Psi''(z)&=&{4\pi\lb n(z)}\label{e7}
\eea
\endnumparts
In addition, the variation with respect to $\Psi_s$ gives the usual electrostatic
boundary condition of the form
\be
  \Psi'(D/2)=\Psi_s^\prime=\frac{4\pi\lb\eta_s}{a^2}\label{e7a}
\ee
The Lagrange multiplier, $\lambda_s$, acts as a chemical potential
with the important difference that it is not related  to any bulk
reservoir, but rather to the concentration at the midplane, $n(0)$.

\begin{figure}
\centering
 \includegraphics[keepaspectratio=true,width=75.3mm,clip=true]{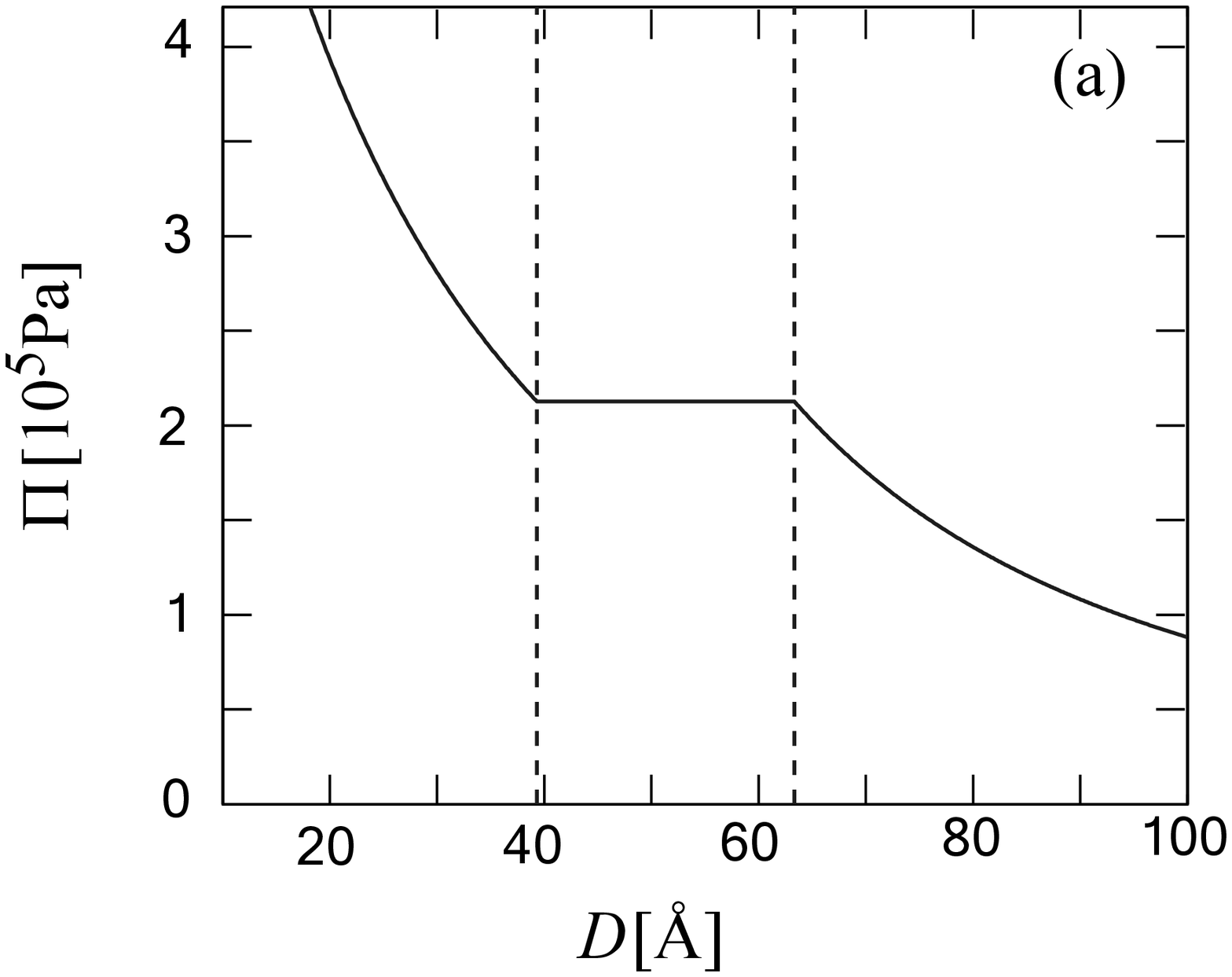}
   \includegraphics[keepaspectratio=true,width=73.6mm,clip=true]{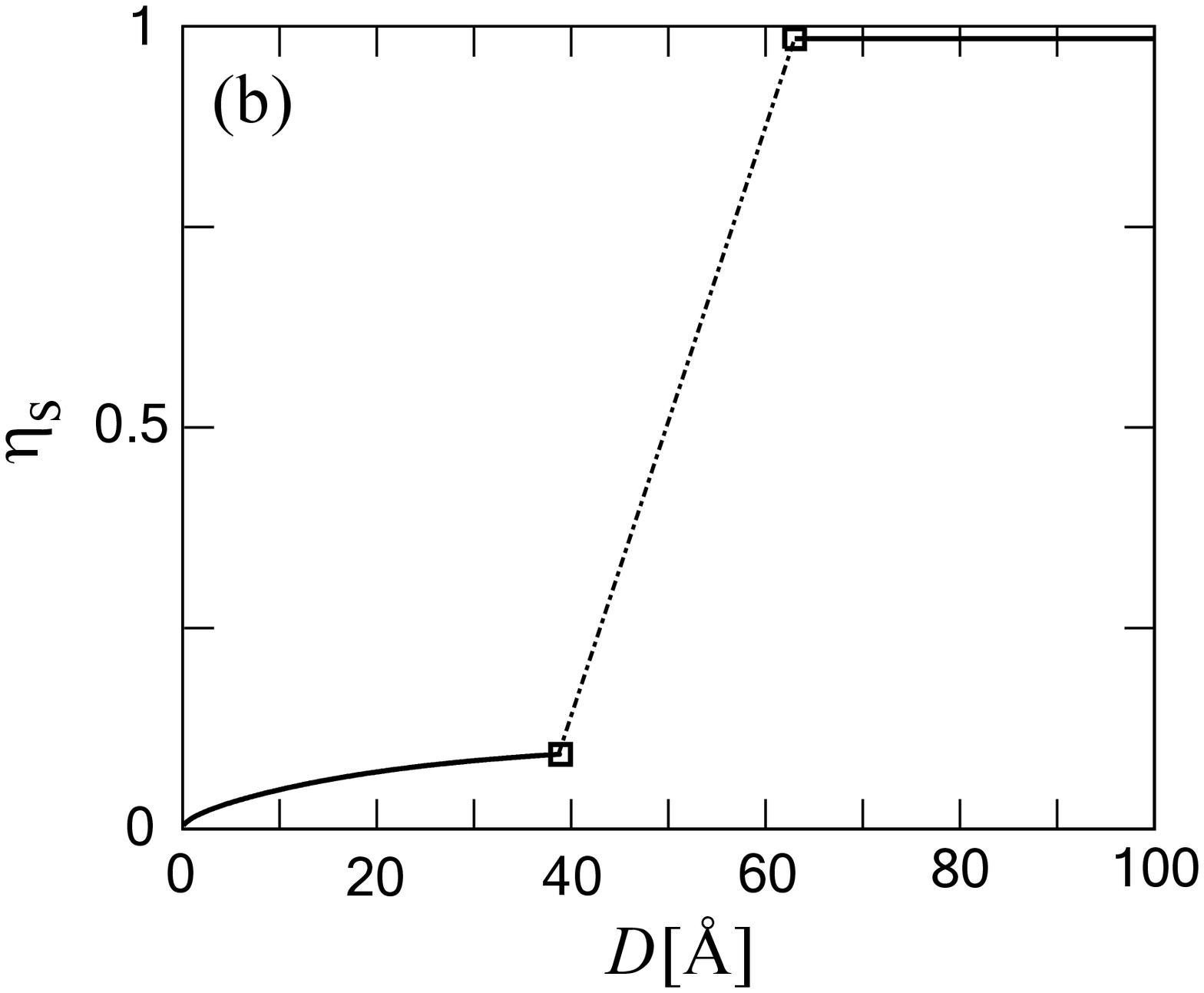}
  \caption{\footnotesize\textsf{(a) The osmotic pressure $\Pi$ in units of $10^5$\,Pascals; and,
  (b) the area fraction $\eta_s=a^2\sigma /e$ of surface charges,
  as function of inter-lamellar spacing $D$ for $\alpha_s=-6$, $\chi_s=12$
  and $a=8$\,\AA. The Maxwell construction
  gives  a coexistence between a phase with $D\simeq 39$\,\AA\ and low $\eta_s\le 0.1$,
  and another with $D\simeq 64$\,\AA\ and  $\eta_s \lesssim 1$. In (b)
  the two coexisting phases are denoted by squares and the dotted-dashed line shows the tie-line
  in the coexisting region.}   \label{3}}
\end{figure}

\begin{figure}
\label{fig4}\centering
\includegraphics[keepaspectratio=true,width=74.11mm,clip=true]{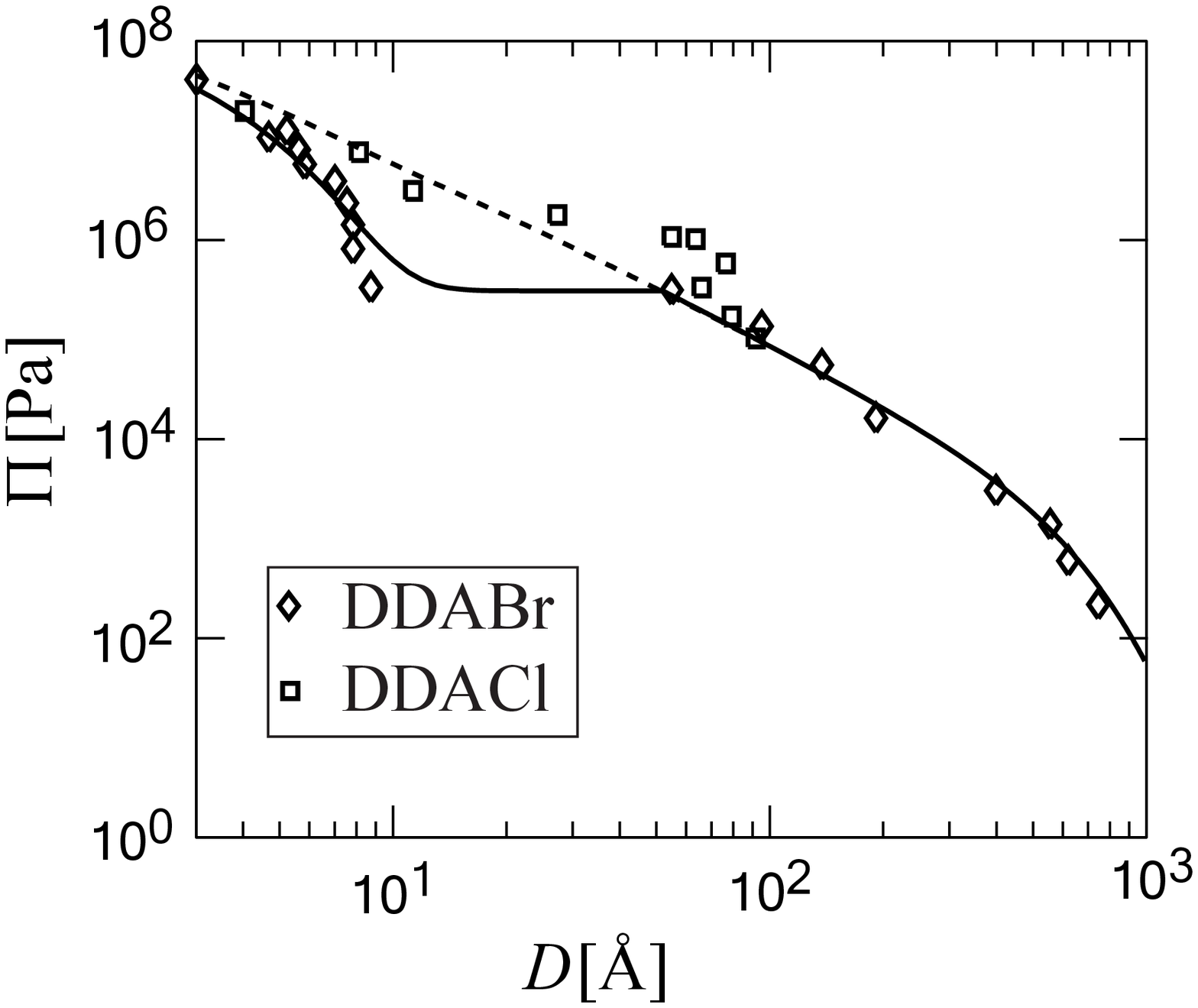}
  \caption{\footnotesize\textsf{Fit to the experimental osmotic pressure isotherm
$\Pi(D)$  of Ref. \cite{zemb98} on a log-log scale.  The diamonds
and squares are the data points for DDABr and DDACl, respectively,
reproduced  from Ref.~\cite{zemb98}.  The solid line is the best fit
of the model to the phase  transition seen for DDABr  with
$\alpha_s=-7.4$, $\chi_s=14.75$ and $a=8$\,\AA .  The fit also
includes a hydration contribution (parameters for the form:
$\Pi_{\rm hyd}=\Pi_0\exp(-D/\lambda_{\rm hyd})$, with typical values:
$\Pi_0=2.37\cdot 10^8$\,Pa and $\lambda_{\rm hyd}=1.51$\,\AA).
This contribution is particularly important at the low $D$
region of the  DDABr isotherm. A small amount of salt is added in
the fits in the experiment  ($n_b=0.5$\,mM). The dashed line is
the fit to the DDACl (no  transition) and all parameters are the
same as for the solid line,  except $\alpha_s=-3.4$. }}
\end{figure}
The non-electrostatic, ion-specific surface  interactions govern the surface
charged surfactant area fraction $\eta_s$, and has the form of a Langmuir-Frumkin-Davis
adsorption isotherm~\cite{davis58}. Combining (\ref{e6}) and
(\ref{e7}) together is
equivalent to the PB equation. Their solution, together with
the adsorption isotherm (\ref{e5}), completely determines the
counterion density profile $n(z)$, the mean electrostatic
potential $\Psi(z)$, and yields the osmotic pressure $\Pi$.
We will apply this formalism to a specific example
of lamellar-lamellar phase transitions next \cite{etay}.

\subsection{The lamellar-lamellar transition in amphiphilic systems}

The typical isotherm $\Pi(D)$ shown in figure \ref{3}a exhibits a first-order phase
transition from one free
energy branch at large inter-lamellar separation $D$ to another at smaller $D$ ---
with a coexistence region in between. For given values of $\alpha_s$ and $\chi_s$
(chosen in the figure to be $\alpha_s=-6$ and $\chi_s=12$), and for large enough
$D$ ($D\ge 64$\AA), most counterions are dissociated from surfaces, $\eta_s\lesssim 1$,
and the osmotic pressure follows the standard  PB theory for (almost) fully dissociated
surfactants. For smaller values of $D$  ($D<39$\AA), most counterions bind to the
surface $\eta_s\le 0.1$ and the isotherm follows another branch, characterized by a
much smaller surface charge of only about $10 \%$ of the fully dissociated value.
In intermediate $D$ range ($39{\rm \AA}\le D\le 64$\AA), the system is in a
two-phase coexistence,  the osmotic pressure $\Pi$ has a plateau and $\eta_s$
changes from one branch to the second (figure \ref{3}b).

Our model is motivated by experiments \cite{zemb98} on the surfactant homolog series:
DDACl, DDABr and DDAI \footnote{DDA stands for
dodecyldimethylammonium and Cl, Br and I correspond to chloride, bromide and iodine,
respectively.}
The main experimental observation is reproduced in figure~4 along with our model fittings.
When Cl$^-$ serves as the counterion, as in DDACl, the osmotic pressure isotherm $\Pi(D)$
follows the usual PB result. When Br$^-$ is the counterion, as in DDABr, one clearly
observes a lamellar-lamellar phase transition from large inter-lamellar spacing of
60\AA to small inter-lamellar spacings of about 10\AA. In addition, for the largest
counterion, I$^-$, as in DDAI, the lamellar stack cannot be swollen to the large
$D$ values branch . We can fit the experimental data by assigning different $\alpha_s$
and $\chi_s$ values to the three homolog surfactants as can be seen in figure~4.
Qualitatively, the different lamellar behavior can be understood in the following way.
The Cl$^-$ counterion is always dissociated from the DDA$^+$ surfactant resulting in
a PB-like behavior, and a continuous $\Pi(D)$ isotherm. For the Br$^-$ counterion,
the dissociation is partial as is explained above (figure~3) leading to a first-order
transition in the isotherm and coexistence between thin and thick lamellar phases.
Finally, for the DDAI, the I$^-$ ion stays associated with the DDA$^+$ surfactant
and there is no repulsive interaction to stabilize the swelling of the stack for
any $\Pi$ value. More details about our model and the fit can be found in \cite{etay}.

Non-electrostatic interactions between counterion-associated and dissociated
surfactants can be responsible for an inplane transition, which, in turn, is
coupled to the bulk transition in the interaction osmotic pressure as can be
clearly seen in figure \ref{3}. This proposed ion-specific interactions are
represented in our model by $\chi_s$ and $\alpha_s$. While at present direct
experimental verification and estimates for the proper $\chi_s$ values are
lacking, the conformational changes induced by the adsorbing ion, together with
van der Waals interaction between adsorbed ions can lead to significant demixing.
Furthermore, because larger ions are expected to perturb the surfactant-water
interface to a larger extent, it is reasonable to expect that the value of
$\chi_s$ will scale roughly with  the strength of surface-ion interactions, $\alpha_s$.

We note that the $\chi_s$ values needed to observe a phase transition,
typically $\approx 10$ (in units of $\kbt$), are quite high \footnote{A detailed discussion about the value of $\chi_s$ is given in \cite{etay}.}. These high values are needed
to overcome the electrostatic repulsion between like-charged amphiphiles,
leading to segregation. The source of this demixing energy, as codified by $\chi_s$, could be
associated with mismatch of the hydrocarbon regions as well as
headgroup-headgroup interactions, such as hydrogen bonding between
neutral lipids, or indeed interactions between lipids across two apposed
bilayers.


\section{Mixed solvents effects in ionic solutions}

So far, in most theoretical studies of interactions between charged
macromolecular surfaces, the surrounding liquid solution was regarded as a
homogeneous structureless dielectric medium within the so-called
``primitive model''. However, in recent experimental studies on
osmotic pressure in solutions composed of two solvents (binary mixture),
the  osmotic pressure was found to be affected by the binary solvent
composition \cite{Rau2006}. It thus seems appropriate to generalize
the PB approach of section~2 by adding local solvent
composition terms to the free energy.

Our approach is to generalize the bulk free energy terms to include regular
solution theory terms for the binary mixture, augmented by the non-electrostatic
interactions between ions and the two solvents in order to account adequately
for the preferential solvation effects \cite{benya}. In this generalized PB
framework the mixture relative composition will create permeability
inhomogeneity that will be incorporated into the electrostatic interactions.

More specifically, our model consists of ions that are immersed in a binary solvent mixture
confined between two planar charged interfaces. Note that
the two surfaces are taken as homogeneous charge surfaces with negative
surface charge $\sigma<0$. We do this to make contact with experiments
on DNA that is also negatively charged \cite{Rau2006}. Though the model is formulated
on a mean-field level it upgrades the regular PB theory in two important
aspects. First, the volume fractions of the two solvents, $\phia$ and
$\phib=1-\phia$, are allowed to vary spatially. Consequently, the dielectric
permeability of the binary mixture is also a function of the spatial coordinates.
In the following, we assume that the local dielectric response $\varepsilon\vecr$
is a (linear) compositionally weighted average of the two permeabilities $\epsa$ and $\epsb$:
\be
\label{linear_eps0}
\varepsilon\vecr=\phia\vecr\epsa+\phib\vecr\epsb\, ,
\ee
or,
\be
\label{linear_eps}
\varepsilon\vecr=\epsz-\phi\vecr\epsr\, ,
\ee
where we define $\phi\equiv\phib$, $\epsz\equiv\epsa$ and $\epsr\equiv\epsa-\epsb$.
This linear interpolation assumption is not only commonly used but
is also supported by experimental evidence \cite{linear_dielectric1}.
Note that the incompressibility condition satisfies $\phia+\phib=1\,$,
meaning that ionic volume fractions are neglected.

The second important modification to the regular PB theory is the
non-electrostatic short-range interactions between ions and solvents.
Those are mostly pronounced at small distances and lead to a reduction
in the osmotic pressure for macromolecular separations of the order 10-20\,\AA.
Furthermore, it leads to a depletion of one of the two solvents from the
charged macromolecules (modeled here as planar interfaces), consistent with
experimental results
on the osmotic pressure of DNA solutions  \cite{Rau2006}.

The model is based on the following decomposition of the free energy
$$F=F_{\rm PB}+F_{\rm mix}+F_{\rm sol}.$$The first term, $F_{\rm PB}$
is the PB free energy of a 1:1 monovalent electrolyte as in (\ref{F_PB})
with one important modification. Instead of a homogeneous permeability,
$\varepsilon$, representing a homogeneous solution, we will use a
spatial-dependent dielectric function $\varepsilon\vecr$ for the binary liquid mixture.
The second term, $F_{\rm mix}[\phi]$, accounts for the free energy
of mixing given by regular solution theory:
\be
\label{FE_bm}
F_{\rm mix}=\frac{\kbt}{a^3}\int\D^3r\,\left[\phi\ln \phi+
(1-\phi)\ln (1-\phi)+\chi\phi(1-\phi) - \frac{\mu_{\phi}}{\kbt}\phi
\right]
\ee
The interaction parameter, $\chi$, is dimensionless (rescaled by $\kbt$).
As the system is in contact with a bulk reservoir, the relative composition
$\phi$ has a chemical potential $\mu_{\rm \phi}$ which is determined by the
bulk composition $\phi_{b}$. For simplicity, we take the same molecular
volume $\sim a^3$ for both A and B components.

The third term, $F_{\rm sol}$,  originates from the preferential non-electrostatic
interaction of the ions with one of the two solvents. We assume that this preference
can be described by a bilinear coupling between the ion densities, $n_\pm$, and the
relative solvent composition $\phi$. This is the lowest order term that accounts
for these interactions. The preferential solvation energy, $F_{\rm sol}$, is then given by
\be\label{FE_s}
F_{\rm sol}=\kbt \int\D^3r\,\left(\alpha_{+} n_{+}+\alpha_{-} n_{-}\right)\phi
\ee
where the dimensionless parameters $\alpha_\pm$ describe the solvation preference of the ions,
defined as the difference between the solute (free) energies dissolved in the A and B
solvents. Finally, to all these bulk terms one must add a surface term, $F_s$,
[as in (\ref{Fs})] describing the electrostatic interactions between charged
solutes and confining charged interfaces.

In thermodynamic equilibrium, the spatial profile of the various degrees of freedom
characterizing the system is again obtained by deriving the appropriate Euler-Lagrange (EL)
equations via a variational principle. The EL equations are then reduced to
four coupled differential equations for the four degrees
of freedom, $\psi\vecr$, $n_{\pm}\vecr$ and $\phi\vecr$. First we have the Poisson equation
\numparts
\be
\label{3d_eom_psi} {\bf \nabla}\cdot\left(
\frac{\varepsilon}{4\pi}{\bf \nabla}\psi\right)+e(n_{+}-n_{-})=0
\ee
then the Boltzmann distribution
\be
\label{3d_eom_n} \pm \frac{e\psi}{\kbt}+\ln(n_{\pm}a^3)+\alpha_{\pm}\phi-\mu_\pm=0
\ee
and finally the EL equation for the density field $\phi\vecr$
\bea
\label{3d_eom_phi}
\ln
\left(\frac{\phi}{1-\phi}\right)&+&\frac{\epsr a^3}{8\pi\kbt}({\bf
\nabla}\psi)^2
 +\chi(1-2\phi)  \nonumber \\
 &+~& a^3\left(\alpha_{+} n_{+} + \alpha_{-} n_{-}\right)
-\frac{\mu_\phi}{\kbt}=0
\eea
\endnumparts
At the charged interface, the electrostatic boundary condition stems from the variation
of $F_s$ with the difference that  $\varepsilon(\phi)$ has a surface value:
$\varepsilon_{s}=\epsz-\epsr\phi_s$, so that the boundary condition becomes
\be
\label{BC}
\hat{{\bf n}}\cdot {\bf \nabla}\psi\bigg{|}_s = -\frac{4\pi e}{\varepsilon_s}\sigma,
\ee
Again the boundary condition states the electroneutrality condition of the
system  as can be shown by the integral form of Gauss' law.

\begin{figure}[t!]
\centering
\includegraphics[keepaspectratio=true,width=78.75mm,clip=true]{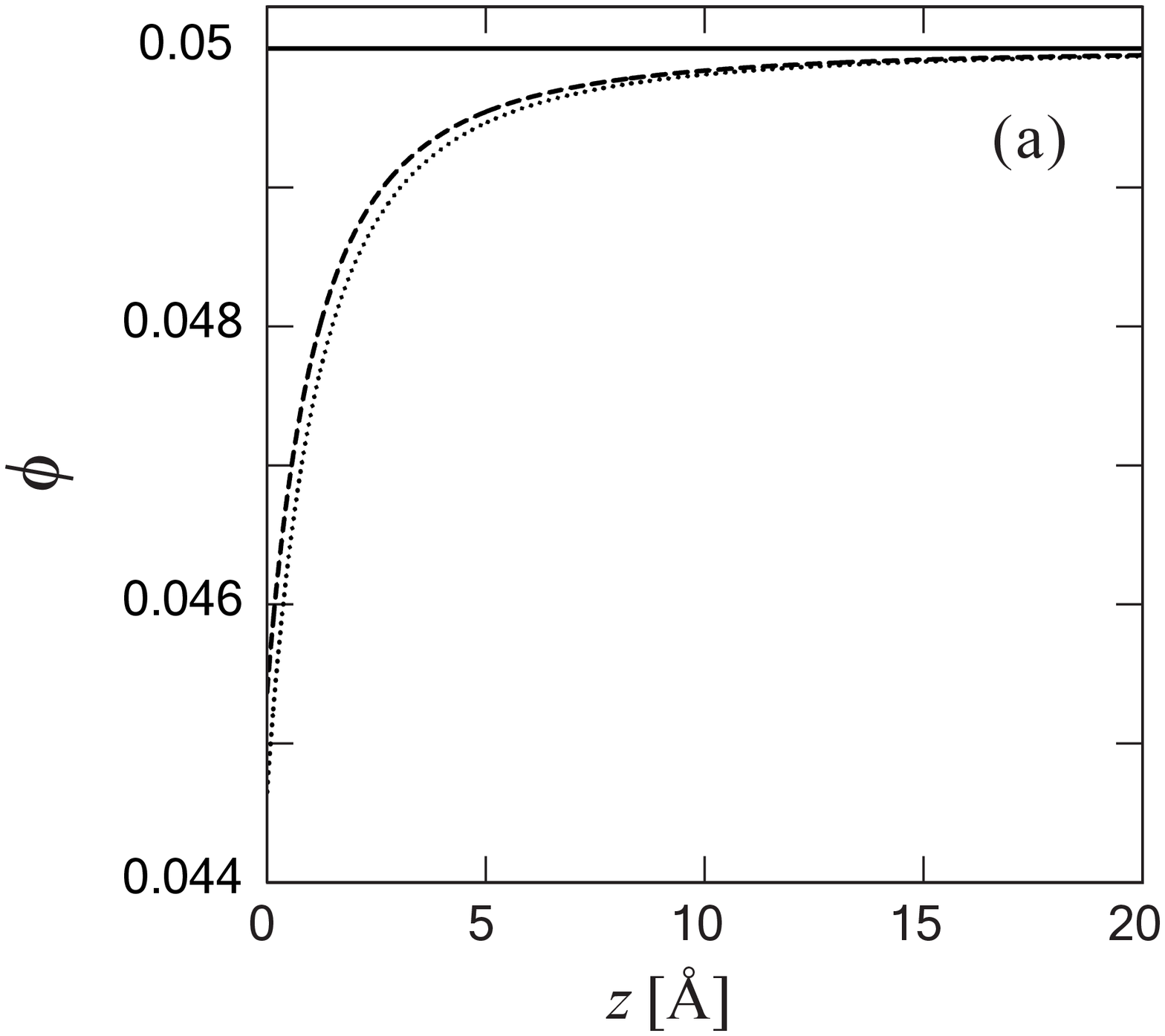}
\hspace{2mm}\includegraphics[keepaspectratio=true,width=74.2125mm,clip=true]{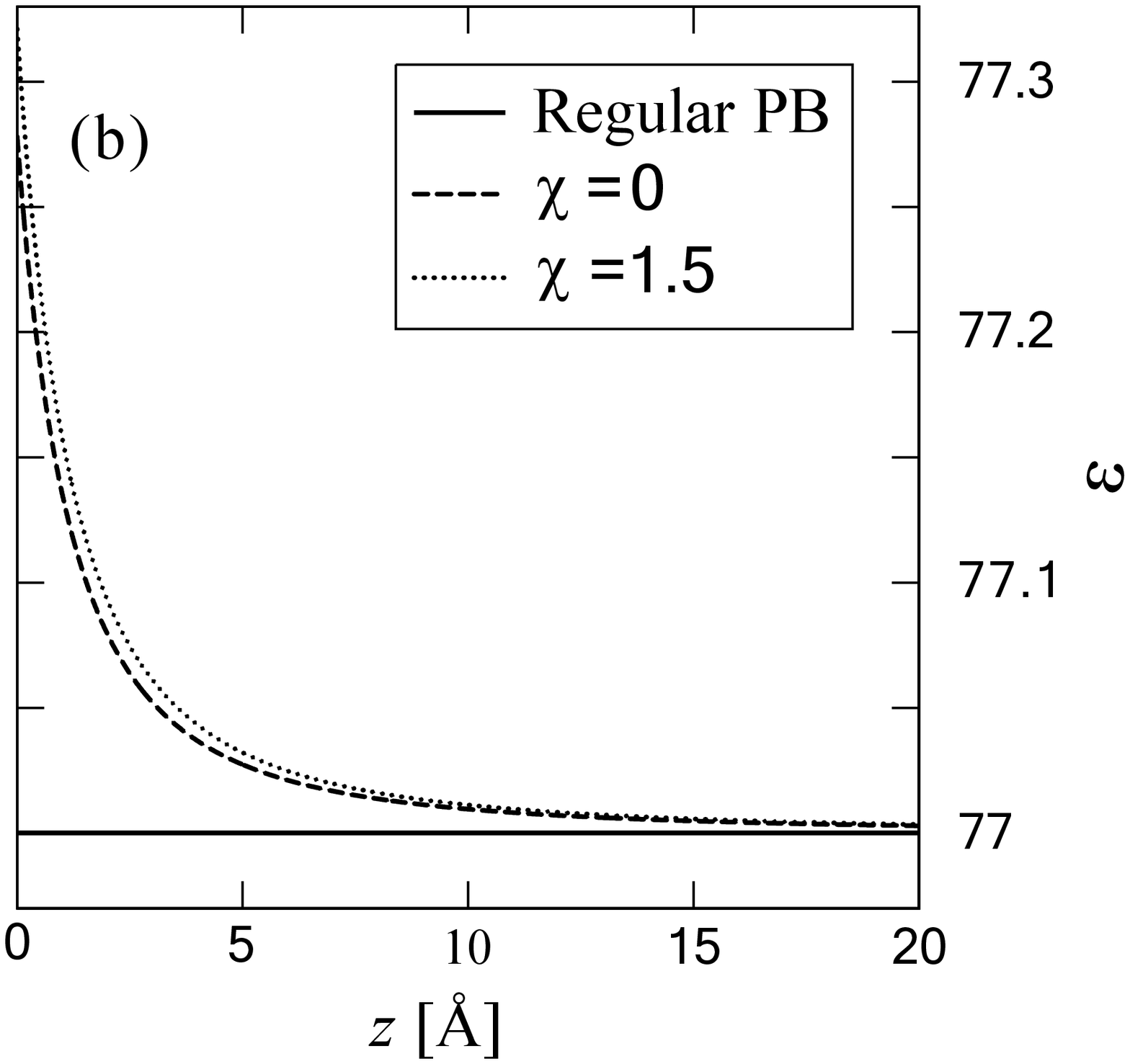}
\caption{\footnotesize\textsf{Spatial profiles of (a) the solvent relative composition
$\phi$ and (b) the permeability $\varepsilon$. The regular PB with
homogeneous dielectric constant $\varepsilon=77$ (solid line) is
compared with our modified PB for binary mixture with and without
short-range interactions, $\chi=0$ (dashed line) and $\chi=1.5$
(dotted line), respectively. Other parameters are:
$\sigma=-1/100$\AA$^{-2}$, $n_{\rm b}=10^{-4}$M, $\varepsilon_\mathrm{A}=80$,
$\varepsilon_\mathrm{B}=20$ and $\phi_{\rm b}=0.05$. In all the cases, no
preferential solvation is considered, $\alpha_\pm=0$.}\label{fig5}}
\end{figure}

By solving the above set of equations, one can obtain the spatial profiles of the various
degrees of freedom at thermodynamic equilibrium. For a general geometry, these equations
can be solved only numerically to obtain spatial profiles for
$\phi$ and $n_\pm$. The osmotic pressure can then be evaluated
via an application of the first integral of (\ref{3d_eom_psi}),
(\ref{3d_eom_n}), (\ref{3d_eom_phi}).This pressure is a function of the
inter-plate separation $D$ and the experimentally controlled
parameters $\alpha_+$, $\phi_b$ and $n_b$.

\begin{figure}[t!]
\centerline{ \includegraphics[keepaspectratio=true,width=70mm,clip=true]{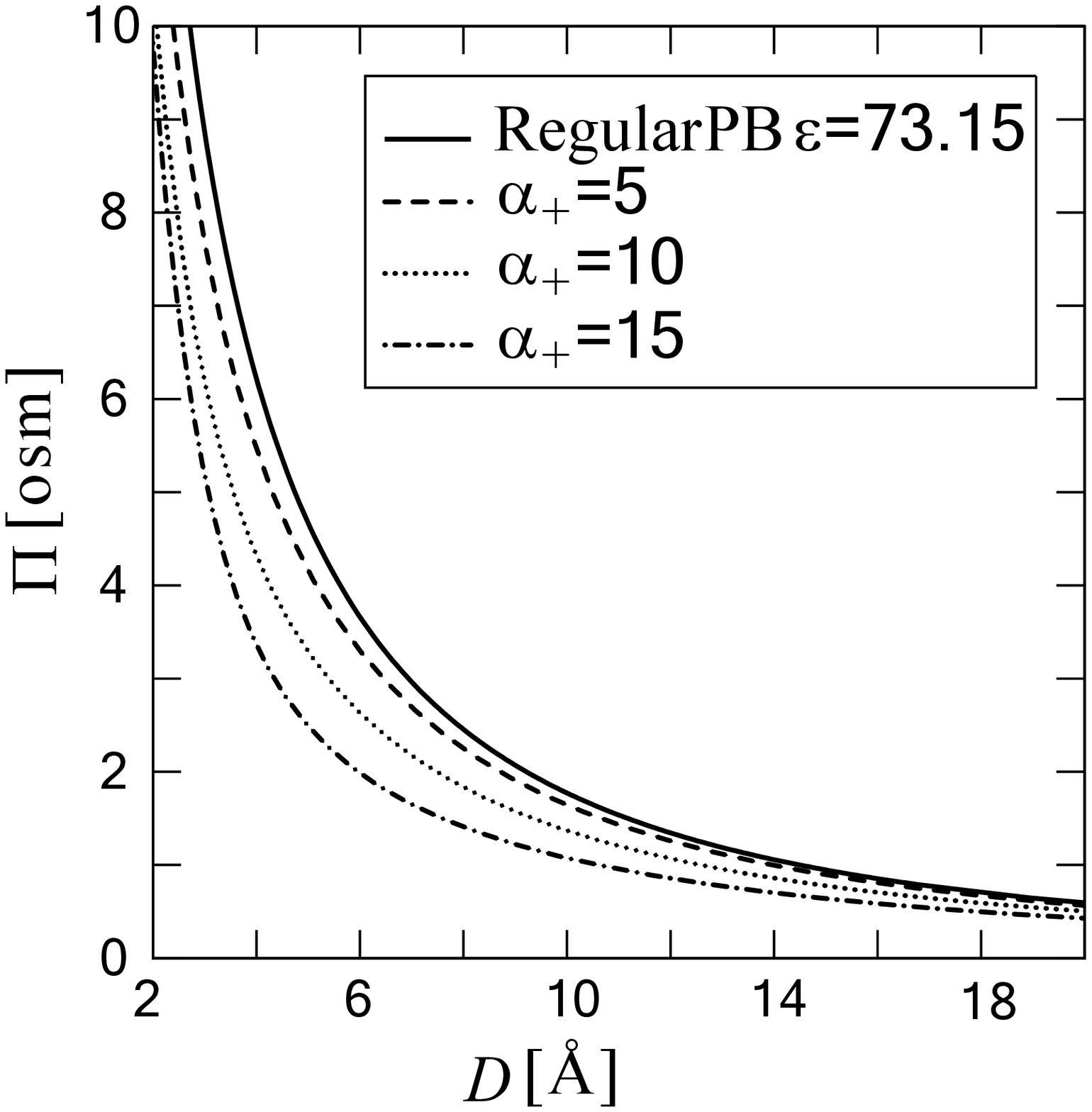} }
\caption{\footnotesize\textsf{The dependence of pressure on separation $D$
for various ion-solvent interaction strengths $\alpha_+$. Other parameters are:
$\sigma=-1/100$\AA$^{-2}$, $n_{\rm b}=10^{-4}$M, $\varepsilon_\mathrm{A}=80$,
$\varepsilon_\mathrm{B}=4$ and $\phi_b = 0.09$.} \label{fig6}}
\end{figure}

For two identically charged planar surfaces, in the absence of preferential solvation,
the numerical solutions of the EL equations show (see figure~\ref{fig5})
that the density profiles  and, therefore, the osmotic pressure undergo only
small modifications. By adding the preferential
solvation term as quantified by $\alpha_+$, one observes a considerable correction
to both the density profile, as well as the pressure (see figure \ref{fig6}).
Most notably is the reduction in the osmotic pressure at small separations
(10-20\,\AA) due to the coupling between ion density and solvent local
composition.

\section{Polarizable ions in solution}

Ion-specific effects as  manifested  through the Hofmeister series
have been recently associated with ionic polarizability, especially
in the way they affect  van der Wails electrodynamic interactions
between  ions and  bounding interfaces \cite{ninham1}. However,
this does not provide the full description, since the
ionic polarizability also modifies the electrostatic interactions
of ions with the surface charges.

In what follows we  generalize
the PB theory in order to include also the contribution of the ionic
polarizability to the overall electrostatic interactions. This inclusion
leads to a new  model that again supersedes the standard PB theory. Note, that
in a more complete and consistent treatment,
the polarizability should also be taken
into account in the electrodynamic van der Wails interactions,
in addition to what we propose here.

We begin along the lines of section 5, while delimiting
ourselves  to the counterion polarizability in the electrostatic
part of the free energy. In order to discuss a manageable set
of parameters, we discard   other possible non-electrostatic terms
in the total free energy and concentrate exclusively on the changes
brought about in the ionic density profile and the interaction osmotic pressure.

The free energy has the same form as $F_{\rm PB}$ for
counterion only case (\ref{F_PB_counter}) with one important
difference that $\varepsilon=\varepsilon(n)$ is now a function
of the counterion density $n$
\be
\label{free_energy}
F/A=\int\D z\left[-\frac{\varepsilon(n)}{8\pi}\psi'^2+e n \psi
+\kbt n(\ln n -1)-\mu n\right]
\ee
The equilibrium equations read
\begin{numparts}
\bea
\label{eqlbrm_eq_1}
\frac{\delta F}{\delta \psi}=4\pi en +\frac{\D}{\D z}\left[\varepsilon(n) \psi'\right]=0 \\
\label{eqlbrm_eq_2}
\frac{\delta F}{\delta n}=-\frac{1}{8\pi}\frac{\partial
 \varepsilon(n)}{\partial n}\psi'^2 +e\psi+\kbt\ln n =\mu
\eea
\end{numparts}
The first equation is a generalization of the Poisson equation and the second is
a generalization of the Boltzmann distribution.
For the free energy (\ref{free_energy}) we can use the general first integral of the system that
gives the pressure in the form:
\be
\label{pressure}
P=-\frac{1}{8\pi}\left(\varepsilon(n)+\frac{\partial \varepsilon(n)}{\partial n}n\right)\psi'^2+\kbt n
\ee
The first term is an appropriately modified form of the
Maxwell stress tensor while the second one is the standard
van 't Hoff term.

>From the first integral  we furthermore derive the following relation:
\be
 \psi'=\sqrt{\frac{8\pi\kbt(n - \tilde{P})}{\varepsilon(n)+
\frac{\partial \varepsilon(n)}{\partial n}n } }
\ee
where $\tilde{P}=P/\kbt$ is the rescaled pressure.
Substituting this relation in (\ref{eqlbrm_eq_1}), we end up with a
first-order ordinary differential equation for the ion density $n$:
\be
\label{n_ode}
\frac{\D n}{\D z}=-\sqrt{\frac{2\pi e^2}{\kbt}}\frac{n}{{\partial f(n)}/{\partial n}}
\ee
where
\be
f(n)=\varepsilon(n)\sqrt{\frac{n-\tilde{P}}{\varepsilon(n)+\frac{\partial \varepsilon(n)}{\partial n}n}}
\ee
is a function of the variable $n$ only. Equation (\ref{n_ode})
can be integrated explicitly either analytically or numerically,
depending on the form of $f(n)$.  Note that in  thermodynamic
equilibrium, the total osmotic pressure $P$ is a constant.

The boundary condition for a constant surface charge is given by
\be
\label{bc}
\varepsilon(n_{\rm s}) \psi'\big|_{\rm s}=4\pi e|\sigma|
\ee
where $\varepsilon(n_s)$ and $n_s$ are the surface
values of the dielectric function and ion density, respectively.
Using the pressure definition, we arrive at an algebraic equation for
the surface ion density:
\be
n_{\rm s}-\tilde{P}=\frac{2\pi e^2\sigma^2}{\kbt\varepsilon^2(n_{\rm s})}
\left({\varepsilon(n_{\rm s}) +\frac{\partial \varepsilon(n_{\rm s})}
{\partial n_{\rm s}}  n_{\rm s}}\right)
\ee
For a single surface the pressure vanishes
$\tilde{P}=0$ (with analogy to two surfaces at infinite separation, $D\to \infty$),
and the basic equations simplify considerably.

As a consistency check, we take the regular case, {\it i.e.} a single
charged surface with a homogeneous dielectric
constant. In this case the function $f(n)$ takes the form $f(n)=
\sqrt{\varepsilon_0 n}$, and (\ref{n_ode}) reads:
\be
\frac{\D n}{\D z}=-\sqrt{8 \pi \l_{\rm B}}n^{\frac{3}{2}}
\ee
which gives the well known Gouy-Chapman result for the
counter-ion profile close to a single charged plate (in absence of added salt) :
\be
\label{n_gc}
n(z)=\frac{1}{2\pi\lb(z+\lgc)^2}
\ee
where $\lgc=1/(2\pi\lb|\sigma|)$ is the Gouy-Chapman length, found by satisfying the boundary condition (\ref{bc})

We assume that to lowest order the dielectric constant can be expanded as a function of the counterion concentration $n$ as:
\be
\varepsilon(n)=\varepsilon_0+ \beta n + {\cal O}(n^2),
\ee
where $\varepsilon_0$ is the
dielectric constant of the solvent and $ \beta={\partial \varepsilon}/
{\partial n}|_0$ is a system parameter describing the molecular
polarizability of the counterions in the dilute counterion limit.

The derivative of $f(n)$ to be used in (\ref{n_ode}) reads:
\be
\frac{\partial f}{\partial n}=\frac{2\beta^2n(n-\tilde{P})+\varepsilon(n)
[\varepsilon(n)+\beta n]}{2\sqrt{(n-\tilde{P})[\varepsilon(n)+\beta n]^3}}
\ee
and (\ref{n_ode}) can now be solved explicitly for $n(z)$.

\begin{figure}\centering
\includegraphics[keepaspectratio=true,width=72mm,clip=true]{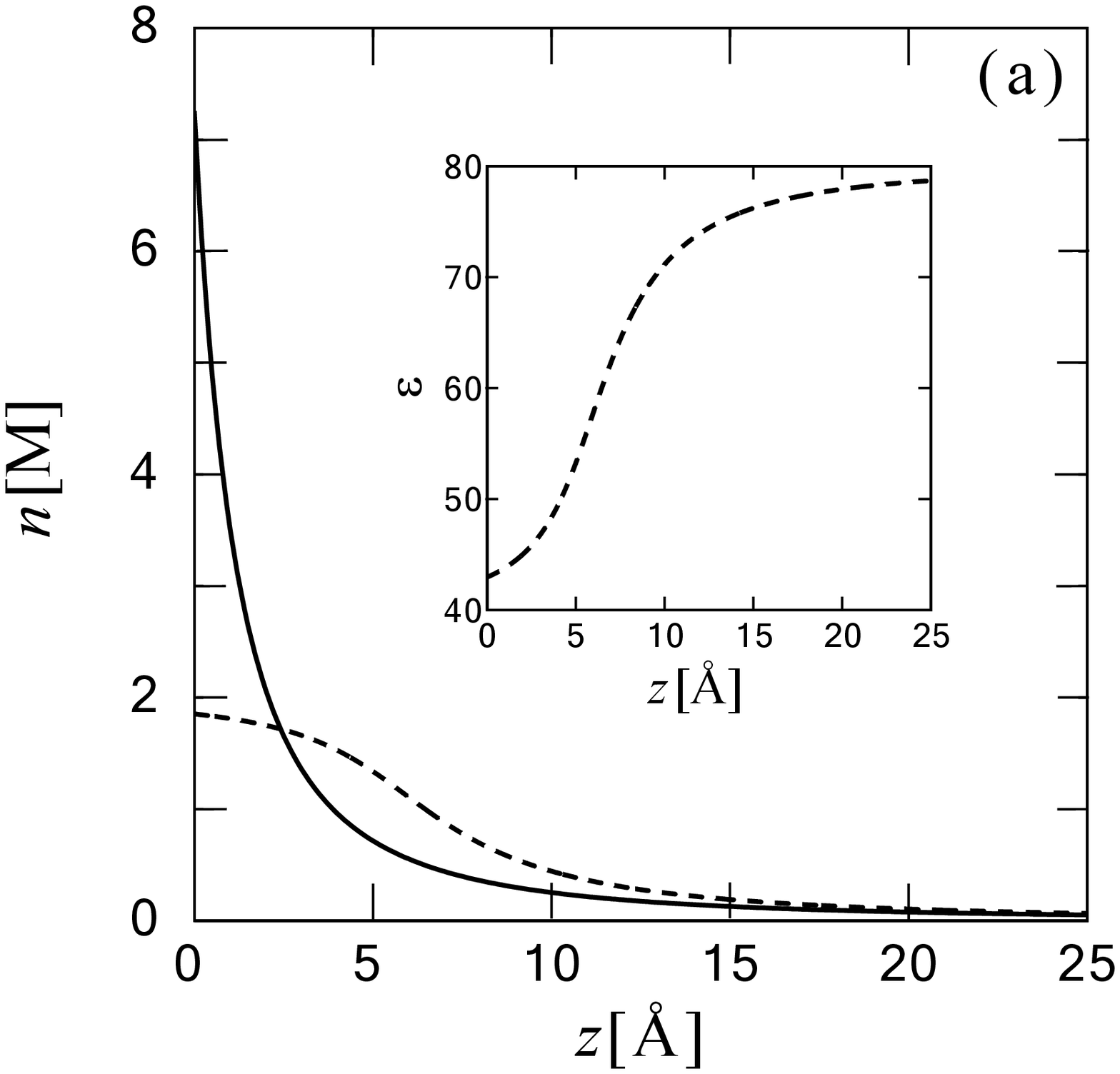}
\includegraphics[keepaspectratio=true,width=69.5mm,clip=true]{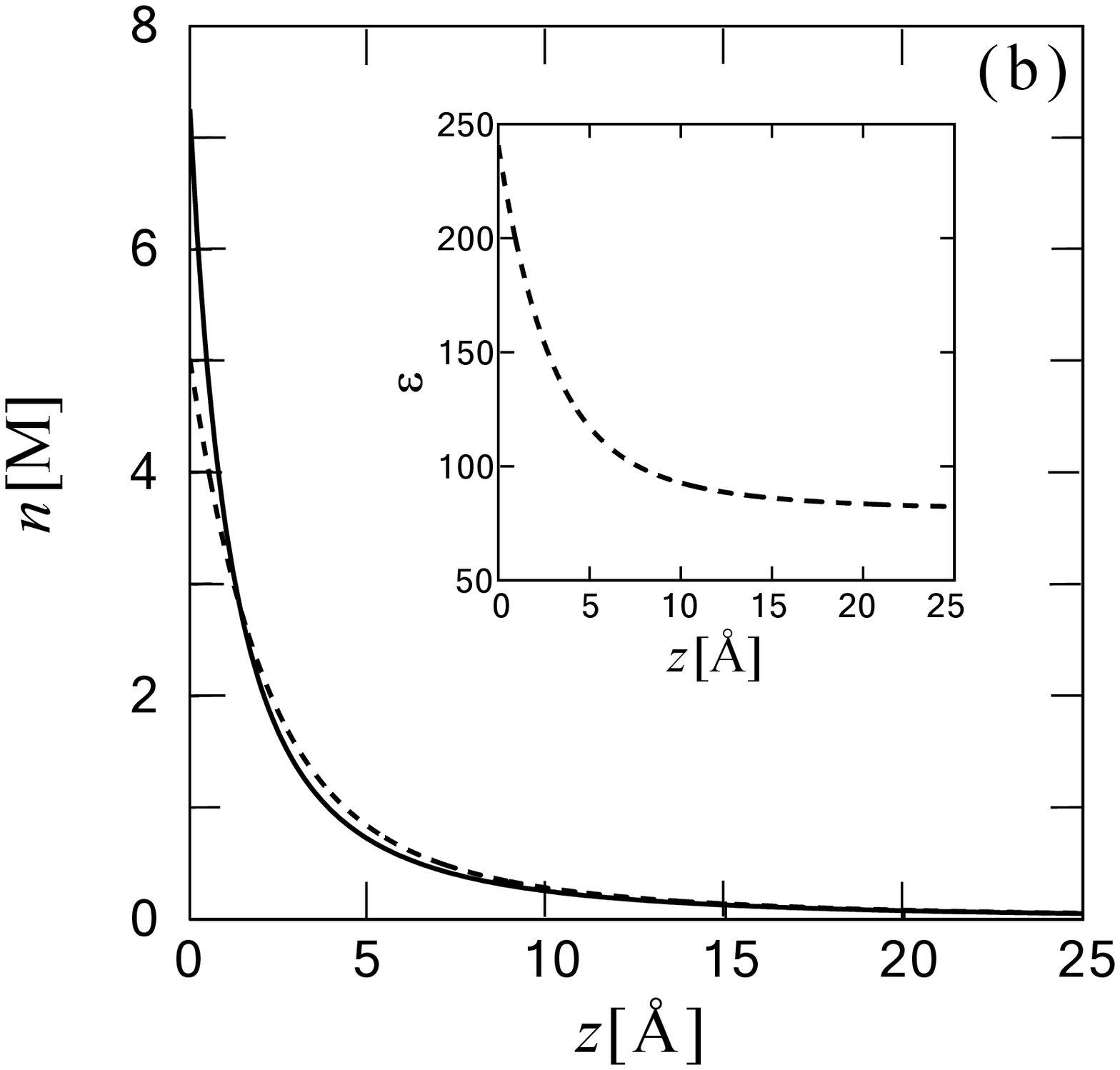}
\caption{\footnotesize\textsf{Ion density and permeability profiles. The
solid lines in both panels represent the case where the ions have no contribution to the
permeability, $\beta=0$, given by (\ref{n_gc}). In (a) the dashed line shows
the ion density where their averaged permeability is smaller than the
solution dielectric constant, $\beta=-30\,$M$^{-1}$. The permeability
of the second case is shown in the inset. In (b) the dashed line shows the ion density
where their averaged permeability is smaller than the solution dielectric
constant, $\beta=20\,$M$^{-1}$. The permeability of the second case is
shown in the inset. Other parameters in both (a) and (b) are:
$\sigma=-1/100\,$\AA$^{-2}$, and $\varepsilon_0=80$.} }\label{fig7}
\end{figure}

In figure~\ref{fig7} we show the spatial profile of the counterion density away
from single charged surface. No extra salt is added and the pressure is zero
$\tilde P = 0$ for a single plate as mentioned above.  In figure~\ref{fig7}a
we present the case where $\beta<0$ ({\it i.e.}, the dielectric permeability
in the regions of high  ion density is smaller than in  the pure solution).
The ion density profile exhibits a somewhat ``flat" behavior, indicating some
saturation in the vicinity of the surface, while at a distances of $\sim 5\,$\AA ~
and further away from the surface the density decays strongly to zero.
In figure~\ref{fig7}b the density profile is shown for the opposite case
where $\beta>0$. Namely, when the dielectric permeability in the regions of
high  ion density is larger  than in  the pure solution. The profile here does
not show large deviations from the usual Gouy-Chapman behavior even close to the surface.

Obviously, the effect of ionic polarizability strongly depends  on the sign of
the ionic polarizability, $\beta$.  It appears that the dependence of the
dielectric constant on ionic density introduces effective interactions
between the ions and the bounding surface. For $\beta <0$, these additional
interactions seem to be strongly repulsive and long ranged. They lead to a
depletion of the ions in the vicinity of the surface. In the opposite case
where for $\beta>0$, the interactions are also repulsive but extremely weak
and the electric double layer structure remains almost unperturbed by the
ionic polarization. This  sets a strong criterion for ion specificity because
the ions can be differentiated according to the
sign of their polarizability affecting their surface attraction.

\section{Conclusions}

We presented here several attempts to generalize the PB theory by including
in the free energy additional terms to the standard electrostatic  and ideal
entropy of  mixing. We also  showed how these terms lead to modifications of the
PB equation and its boundary conditions. More specifically, we have aimed to include
additional interactions between dissolved ions, such as finite ion size and
polarizability,  as well as their solvation and interactions with bounding surfaces.
All these endeavors are done within the mean-field approximation while neglecting
charge density fluctuations and ion correlations.

The approach presented here  amends  the free energy in specific ways giving rise
to modified electric bilayer charge distribution and  ensuing interactions between
charged surfaces as mediated by ionic solutions.  Some modification wrought by the
non-electrostatic terms can be interpreted in retrospect as specific interactions
of the ions with the bounding surfaces or between the ions themselves.

To verify these models, results should be compared with appropriate experiments.
In some cases,
such comparisons are feasible and are qualitatively favorable~\cite{etay,benya}.
On the other hand, comparisons with extensive all-atom MC or MD simulations are not obvious since
these simulations also contain multiple parameters that have no obvious analogue.
It would, nevertheless, be valuable to link these approaches with appropriately
designed experiments or even to check the predictions of these approaches in realistic systems.

\ack
We thank Y. Marcus for helpful discussions. One of us (DA) acknowledges support
from the Israel Science Foundation (ISF) under grant no. 231/08 and the US-Israel
Binational Foundation (BSF) under grant no. 2006/055. RP and DH would like to
acknowledge the support from the Israeli and Slovenian Ministries of Science
through a joint Slovenian-Israeli research grant. The Fritz Haber research
center is supported by the Minerva foundation, Munich, Germany.

\section*{References}


\begin{thebibliography}{99}

\bibitem{Andelman} Poon W C K and Andelman D 2006 {\it Soft Condensed Matter Physics
in Molecular and Cell Biology}, (Taylor \& Francis).

\bibitem{Oosawa} Oosawa F 1968 {\it Biopolymers} {\bf 6} 1633

\bibitem{Naji} Naji A, Jungblut S, Moreira A G and Netz R R 2005 {\it Physica A} {\bf 352} 131

\bibitem{hoda} Boroudjerdi H, Kim Y W, Naji A, Netz R R, Schlagberger X and Serr A 2005 {\it Phys. Rep.} {\bf 416} 129

\bibitem{hydration} Chan D Y C, Mitchell D J, Ninham B W and
Pailthorpe B A 1979 in {\it Water: A Comprehensive Treatise} (Vol. 6 Recent Advances) Ed F Franks
(New York: Plenum). Conway B E 1981 {\it Ionic Hydration in Chemistry and Biophysics},
(Studies in Physical \& Theoretical Chemistry) (Amsterdam: Elsevier). Marcus Y 1997 {\it Ion Properties},
CRC; 1st edition. Leikin S, Parsegian V A, Rau D C and Rand R P 1993 {\it Annu. Rev. Phys. Chem.} {\bf  44}
369. Ben-Naim A 1987 {\it Solvation Thermodynamics} (New York: Plenum)

\bibitem{hydrophobic}  Tanford C 1980 {\it The Hydrophobic Effect: Formation of
Micelles \& Biological Membranes}, (Wiley, New York). Chandler D 2005
{\it Nature} {\bf 437} 640. Meyer E E,  Rosenberg K J and Israelachvili J 2006 {\it PNAS} {\bf 103} 15739.
Dill K A, Truskett T M, Vlachy V  and Hribar-Lee B 2005 {\it Annu. Rev. Biophys. Biomol. Struct.} {\bf 34} 173

\bibitem{marcus}
Marcus Y 2007 {\it Chem. Rev.} {\bf 107} 3880. Kalidas C, Hefter G and Marcus Y 2000
{\it Chem. Rev.} {\bf 100} 819

\bibitem{cviklik}
Vlachy N, Jagoda-Cwiklik B, V\'{a}cha R, Touraud D, Jungwirth P and Kunz W 2009
{\it Adv. Coll. Interf. Sci.} {\bf 146} 42

\bibitem{onuki}
Onuki A and Kitamura H 2004 {\it J. Chem. Phys.} {\bf 121} 3143.  Onuki A 2006 {\it Phys. Rev. E}
{\bf 73} 021506. Onuki A 2008 {\it J. Chem. Phys.} {\bf 128} 224704.
Tsori Y and Leibler L 2007 {\it Proc. Natl. Acad. Sci. (USA)}  {\bf 104} 7348

\bibitem{currentninham}
Hofmeister F 1888 {\it Archiv. Exp. Path. Pharm.} {\bf 24} 247. Collins K and Washbaugh
M 1985 {\it Q. Rev. Biophys.} {\bf 18} 323. Harries D and R\"{o}sgen J 2008 {\it Meth. Cell Biol.} {\bf 84} 679

\bibitem{chargeregulation}
Ninham B W and Parsegian V A 1971 {\it J. Theor. Biol.} {\bf 31}
405. Chan D, Perram J W, White L R and Healy T H 1975 {\it J. Chem. Soc., Faraday Trans. 1} {\bf 71} 1046. Chan D,
Healy T and White L R 1976 {\it ibid} {\bf 72} 2844. Prieve D C and Ruckenstein E 1976
{\it J. Theor. Biol.} {\bf  56} 205. Pericet-Camara R, Papastavrou G, Behrens S H and Borkovec M 2004
{\it J. Phys. Chem. B} {\bf 108} 19467.
Von Gr\"unberg H H 1999  {\it J. Colloid Interface Sci.} {\bf 219} 339

\bibitem{evans}
Evans D F and Wennerstroem H 1994 {\it The Colloidal Domain} (Wiley-VCH)

\bibitem{eigen}
Eigen M and Wicke E 1954 {\it J. Phys. Chem.} {\bf 58} 702

\bibitem{iglic}
Kralj-Iglic V and Iglic A 1996 {\it J. Phys. II (France)} {\bf 6} 477

\bibitem{Borukhov}
Borukhov I, Andelman D and Orland H 1997 {\it Phys. Rev. Lett.}
{\bf 79} 435. Borukhov I, Andelman D and Orland H 2000 {\it Electrochim. Acta} {\bf 46} 221.

\bibitem{tresset} Tresset G 2008
{\it Phys. Rev. E } {\bf 78} 061506

\bibitem{israelachvili}
Israelachvili J N 1992 {\it Intermolecular and Surface
Forces} (Elsevier)

\bibitem{messina}
Messina R 2009 {\it J. Phys.: Condens. Matter} {\bf 21} 113102

\bibitem{hill}
Hill T L 1986
{\it An Introduction to Statistical Theromodynamics} (New York: Dover)

\bibitem{pbapproach}
Stern O 1924 {\it Z. Elektrochem.} {\bf 30} 508.
Henderson D 1983 {\it Prog. Surf. Sci.} 13 197. Volkov A G, Deamer D W,
Tanelian D L and Mirkin V S 1997 {\it Prog. Surf. Sci.} {\bf 53} 1. Kjellander R and Marcelja S
1986 {\it J. Phys. Chem.} {\bf 90} 1230.
Kjellander R, Akesson T, Jonsson B and Marcelja S 1992 {\it J. Chem. Phys.} {\bf 97} 1424


\bibitem{zemb98}
Dubois M, Zemb Th, Fuller N, Rand R P and Parsegian V A 1998 {\it J. Chem. Phys.} {\bf 108} 7855

\bibitem{rydall92}
Rydall J R and Macdonald P M 1992 {\it Biochemistry} {\bf 31} 1092

\bibitem{sachs03}
Sachs J N and Woolf T B 2003 {\it J. Am. Chem. Soc.} {\bf 125} 874

\bibitem{etay}
Harries D, Podgornik R, Parsegian V A, Mar-Or E and Andelman D 2006
{\it J. Chem. Phys.} {\bf 124} 224702

\bibitem{davis58}
Davis J T 1958 {\it Proc. Royal Soc. A} {\bf 245} 417

\bibitem{Rau2006}
Stanely R and Rau D C 2006 {\it Bipophys. J.} {\bf 91} 912

\bibitem{benya}
Ben-Yaakov D, Andelman D, Harries D and
Podgornik R 2009 {\it J. Phys. Chem. B} {\bf 113} 6001

\bibitem{linear_dielectric1}
Arakawa T and Timasheff S N,
1985 {\it Methods Enzymol.} {\bf 114} 49

\bibitem{ninham1}
Ninham B W and Yaminsky V 1997 {\it Langmuir}
{\bf 13} 2097. Bostrom M, Williams D R M and Ninham
B W 2001 {\it Phys. Rev. Lett.} {\bf 87} 168103

\bibitem{ninham2}
Kunz W,  Nostro P and Ninham B W 2004
{\it Curr. Opin. Colloid Interf. Sci.} {\bf 9} 1

\bibitem{ajdari}
Kilic M S, Bazant M Z and Ajdari A 2007
{\it Phys. Rev. E} {\bf 75} 021502

\bibitem{blume00}
Garidel P, Johann C and Blume A 2000
{\it J. Lip. Res.} {\bf 10} 131

\end{thebibliography}
\end{document}